# The influence of dark excitons on the electroabsorption spectrum of polyacetylene

Jaspal Singh Bola[1]*, Ryan M. Stolley[2]*, Prashanna Poudel[1], Joel S. Miller[2], Christoph Boheme[1] and Z. Valy Vardeny[1]**

[1] Department of Physics & Astronomy, University of Utah, Salt Lake City, Utah 84112
[2] Department of Chemistry, University of Utah, Salt Lake City, Utah 84112

Abstract

This study revisits the electroabsorption (EA) spectrum of polyacetylene, $(CH)_x$ thin-films, for both the *cis*- and *trans*-isomers, as functions of the electric field strength, isomerization degree, and light polarization states. The EA spectrum of *cis*-$(CH)_x$ reveals an oscillatory feature that follows the Stark shift-related first derivative of the material's absorption spectrum that contains $v$(0-1) and $v$(0-2) sidebands of the excited C=C stretching vibration that agrees well with the Raman scattering (RS) spectrum. In contrast, the EA spectrum of *trans*-$(CH)_x$ does not match the first derivative of the material's absorption spectrum, and the phonon sideband frequency does not agree with the RS spectrum. In addition, the EA spectrum of *trans*-$(CH)_x$ reveals a band below the first allowed $1B_u$ exciton. We interpret this feature as due to the electric field activated even-parity dark (forbidden) exciton, namely $mA_g$ (m>1), showing that the 'nonluminescent' *trans*-$(CH)_x$ is due to the reverse order of the excited states, where a dark $mA_g$ exciton lies below the allowed $1B_u$ exciton. This agrees with the unusual phonon sideband in *trans*-$(CH)_x$ absorption, since the excited state attenuation caused by the fast internal conversion from $1B_u$ to $mA_g$ influences the apparent frequency that determines the phonon sideband. Consequently, from the EA and RS spectra we estimate the $1B_u$ lifetime in *trans*-$(CH)_x$ to be ~30 fs. Moreover, the integrated EA spectrum of *trans*-$(CH)_x$ shows a traditional Huang-Rhys type series with a relaxation parameter, S ~ 0.5. This indicates that the EA spectrum of the *trans* isomer is also determined by a Stark shift related to the first derivative of the absorption spectrum, but preferentially for the longest chains in the film's chain lengths distribution. This is due to the $N^3$ response of the non-linear susceptibility, $\chi^{(3)}$ (~EA), dependence on the chain length having N monomers.

* These authors contributed equally to this work

** Correspondent author, e-mail val@physics.utah.edu

**Introduction**

Polyacetylene, (CH)$_x$, is composed of chains of carbon atoms having alternating single and double bonds that are bonded to hydrogen [see *Fig*. 1(a)] [1]. The double bonds exist in either *cis* or *trans* arrangements. By varying the reaction temperature, it is feasible to preferentially synthesize each isomer, namely, *cis*-polyacetylene [*cis*-(CH)$_x$] or *trans*-polyacetylene [*trans*-(CH)$_x$] [2-4]. At different synthesis temperatures, both isomers can exist simultaneously, and *cis*-(CH)$_x$ can be transformed into the more stable *trans* configuration through heat treatment. The (CH)$_x$ chains can be as long as ~100 carbon atoms, but there is a distribution of chain lengths with various monomeric units, N. When synthesized, the (CH)$_x$ chains form 'bundles' in the form of fibrils, which can be as long as tens of microns (see *Fig*. 1(b)) [4].

The *trans*-(CH)$_x$ is structurally the simplest π-conjugated polymer (PCP), with two degenerate ground states, or "A" and "B" phases, depending on the 'order direction' of the single and double bonds in the chain [see *Fig*. 1(a)]. The misfit between the A and B phases within a chain forms a domain wall and a nonlinear shape-preserving excitation that is described by the Su, Schrieffer, and Heeger (SSH) model as a mobile soliton [5,6], which may be neutral, $S^0$, or charged, $S^\pm$. Importantly, a neutral soliton has a single electron, and, hence, a spin $s = ½$, whereas positively and negatively charged solitons have zero ($S^+$) and two electrons ($S^-$), respectively, which both are spinless [6-8]. This leads to the so-called excitation having 'reverse spin-charge relationship', in which the charged species is spinless, whereas the spin ($s = ½$) carrying species is charge-neutral. In particular, this model [5], which includes electron-phonon interaction, but neglects electron-electron interaction, predicts that upon photon absorption, the photoexcited electron and hole pair is unstable, thereby relaxing within 100 fs to a charge soliton-antisoliton ($S^+S^-$) pair with a state in the middle of the optical gap. Within this model the nonluminescent nature of *trans*-(CH)$_x$ was taken as evidence that this process indeed dominates the photophysics of the *trans*-(CH)$_x$ isomer [9]. Little attention, however, has been devoted to the *cis*-(CH)$_x$ isomer since it has a non-degenerate structure and shows photoluminescence [9], albeit weak compared to other PCPs [10].

Over the years, the SSH model has attracted substantial attention, where both experimentalists and theoreticians have tried to prove or disprove the photogenerated $S^+S^-$ prediction [6,11-14]. With time, however, other PCPs have been synthesized in which the photophysics is dominated by singlet and triplet excitons [11]. In a typical PCP, such as polyfluorene, polythiophene, and poly(p-phenylene-vinylene) derivatives, the lowest singlet exciton binding energy, namely the $1B_u$ state, is of the order of 0.5 eV [15]. The exciton being the primary photoexcitation species in these polymers indicates that the SSH model is not suitable to describe the photophysics of most PCPs. Even for *trans*-$(CH)_x$, during the last three decades, the study of photoexcitations has revealed several unexpected phenomena, including photogeneration of a neutral excitation, which were not predicted by the SSH model [12,13]. This indicates that the nature of the primary photoexcitations in *trans*-$(CH)_x$ may be very different from the $S^+S^-$ pair predicted by the SSH model. It is therefore expected that the electronic energy levels of *trans*-$(CH)_x$ would be also very different than that predicted by the SSH model.

An effective method to investigate the electronic energy levels in PCPs has been electroabsorption (EA) spectroscopy [16-20]. Using this technique, it is expected that the allowed excitons in PCPs, namely the $B_u$ odd parity excitons, undergo a Stark shift which results in EA spectral features that resemble the first-derivative of the absorption spectrum [21-22]. Also, the forbidden (or dark) excitons with the $A_g$ even parity would become partially allowed due to the symmetry breaking imposed by the external electric field [22]. EA spectroscopy is a very useful method to study whether the $1B_u$ exciton is indeed the lowest exciton in a specific PCP [23], or alternatively there is an $A_g$ state ($mA_g$, where m > 1 is an integer) below it. Kasha's rule states that photoluminescence (PL) at steady state is mainly due to the lowest energy excited state of the material. Accordingly, if the $1B_u$ is indeed the lowest exciton, then the PCP is luminescent; however, if $E(1B_u)>E(2A_g)$, then the specific PCP is non-luminescent. Since *trans*-$(CH)_x$ is non-luminescent, whereas *cis*-$(CH)_x$ is luminescent, it is important to revisit EA spectroscopy of $(CH)_x$ [20, 24-25], in order to corroborate that this material can be described by the same theoretical framework as many other PCPs, without invoking an ultrafast generation of $S^+S^-$ excitation to explain the apparent lack of PL in this polymer. This is particularly important as, upon deposition of a $(CH)_x$ film, the materials contains both *trans* and *cis* isomers, which are non-luminescent and luminescent, respectively.

In the study presented here, we report the synthesis and investigation of $(CH)_x$ films using optical absorption, EA and Raman scattering (RS) spectroscopies, at different stages of the $(CH)_x$ isomerization, with the goal to scrutinize the EA spectra of the luminescent *cis*-$(CH)_x$ and non-luminescent *trans*-$(CH)_x$ for the first-derivative of the material's absorption spectrum.

**Experimental Procedures**

The procedure employed for synthesizing the polyacetylene $(CH)_x$ thin films in this study was a modified version of the method reported by Shirakawa and co-workers in 1999 [26], as detailed in the Supplementary Information (S.I.). The synthesis of the polymerization catalyst was prepared using a freshly prepared catalyst solution (0.003 M) synthesized by the addition of tetra-*n*-butoxytitanium(IV) (Thermo-Fisher), $Ti(OBu)_4$ (251 μL, 0.74 mmol, 1 equiv), to a 1 M solution of triethylaluminium $AlEt_3$ (Sigma Aldrich) in hexanes (2.95 mL, 2.95 mmol, 4.0 equiv) under Schlenk conditions. The resulting solution was heated to 150 °C for 1 h, then allowed to cool to room temperature affording a 0.1 M solution. The catalyst solution was then dissolved in dry, degassed toluene (Thermo-Fisher) to the appropriate polymerization concentration. The polymerization process utilized high purity acetylene gas dissolved in acetone that was further purified by washing through subsequent stages of saturated sodium bisulfite, concentrated sulfuric acid, and passage through packed columns of activated alumina and Drierite®. Thin films of $(CH)_x$ having various thicknesses were deposited on different substrates such as quartz, sapphire, and KBr within a controlled nitrogen and/or acetylene environment. Each substrate was dried in an oven prior to deposition and placed in a pre-dried Schlenk flask equipped with a rubber septum and attached to an appropriately outfitted manifold. Under a nitrogen atmosphere, the catalyst solution was deposited onto the substrate and the atmosphere exchanged for acetylene. Upon exchanging the atmosphere, an immediate formation of the purple polymerization product was apparent. Upon deposition of the desired thickness, the atmosphere was evacuated and the substrate was dried under vacuum. The films were then washed under nitrogen with toluene (3 x 1-2 mL) and dried thoroughly under vacuum. The film thicknesses were then determined by using the well-known optical density strength [27], which turned out to be in a range of ~100 nm to 1 μm. Isomerization from the as-synthesized film by converting the *cis*-$(CH)_x$ chains in the films into *trans*-$(CH)_x$ chains was done by heating the film while being sealed in a nitrogen environment at a temperature of 150 °C, and various durations (~tens of minutes), depending on the desired *cis*-$(CH)_x$/*trans*-$(CH)_x$ ratio.

The optical absorption spectra for the deposited thin films were measured using an in-house set up containing a ¼ met. monochromator with a tungsten lamp that serves as the incident light source directed onto the sample. The thin $(CH)_x$ film was placed in a cryostat capable of maintaining temperatures between 10 to 300 K. The transmitted light intensity was measured using a silicon detector. The Raman scattering spectra were obtained in reflection mode using 488 nm laser at low intensity at room temperature, using a ¼ met. monochromator with spectra measured from 400 to 1900 cm$^{-1}$ with a resolution of 2 cm$^{-1}$.

The EA measurements were conducted on various thin $(CH)_x$ films having different *cis*-$(CH)_x$/*trans*-$(CH)_x$ ratios, deposited on sapphire substrates with patterned metallic, interdigited electrodes, consisting of several hundred 10-mm-wide gold strips [19]. The EA measurement setup is shown in (S.I.) *Fig*. S1. The devices were placed in a cryostat for low temperature measurements. By applying a potential, $V$, to the electrodes, in-plane electric field on the order of $F \approx 10^5$ V/cm, were generated. Typically, a voltage $V = 300$ V was applied at a modulation frequency, $f = 1$ kHz. To probe the EA spectrum, we used an incandescent light source from a tungsten lamp, which was dispersed through a ¼ meter monochromator, focused on the sample, and detected by a silicon photodiode. For the EA experiments, we measured the changes, $\Delta T$, in the transmission spectrum, $T$, using a lock-in amplifier, set to twice the frequency ($2f$) of the applied modulation frequency, and verified that no EA signal was observed at $f$ or $3f$ (see S.I. *Fig*. S2(a)). The spectra of $\Delta T$ and $T$ were measured separately, using the same set-up, and the EA spectrum was subsequently obtained from the spectral ratio, $\Delta T/T$.

**Results and discussion**

The optical absorption spectra of both an 'as-synthesized' $(CH)_x$ film and another film that underwent annealing at 150 °C for 30 min, where the *cis*- to *trans*-$(CH)_x$ isomerization occurs are shown in *Fig*. 1(c). The absorption spectrum of the as-synthesized film shows two superimposed bands having onset at 1.4 and 2.0 eV, respectively. Upon isomerization from *cis* to *trans*, the higher energy band weakens, and thus it is identified as the absorption band of *cis*-$(CH)_x$ with an optical gap of ~2 eV. This is in agreement with the PL spectrum of *cis*-$(CH)_x$ that was measured to peak at 1.97 eV [27]. Consequently, the low energy absorption band is due to the *trans*-$(CH)_x$ isomer in the film with an optical gap at ~ 1.4 eV depending on the *trans/cis* ratio that is consistent with an extremely weak polarized PL band at 1.41 eV having phonon side band at 1.37 eV [14]. Note that each absorption band also contains several superimposed weaker features. These are phonon sidebands of the lowest lying allowed exciton in these two isomers, which become more visible in the EA spectra (see below). The broad absorption band of the *trans* isomer indicates that the *trans* chains in the film have a broad distribution of chain lengths. This reflects that the energy, E(N), of lowest allowed $1B_u$ exciton of the *trans* chains (i.e. polyenes) that depends on the chain length, N, as $E(N) = E_{min} + a/N$ where *a* is a constant and $E_{min}$ is the exciton energy of the longest chain length [28-29]. This broadens the phonon sidebands, as noted in the absorption spectra. Another contribution to the broad absorption band related to *trans*-$(CH)_x$ isomer may be attributed to scattering from the $(CH)_x$ fibrils [24] *Fig*. 1(b).

The resonant Raman scattering (RRS) of the as-synthesized film and the 'annealed' $(CH)_x$ film are shown in *Fig*. 1(d). The RRS spectrum of as-synthesized film contains three resonantly enhanced Raman bands at 909, 1251, and 1539 cm$^{-1}$, which diminish upon isomerization. We assign this trio of Raman bands to the C-C stretching, C-C bending and C=C stretching modes, respectively, of *cis*-$(CH)_x$, in agreement with the literature [30]. The other three bands in the RRS spectrum of the as-synthesized film are observed at 1118, 1290, and 1496 cm$^{-1}$ and do not disappear upon isomerization; thus, they are also assigned to the C-C stretching, C-C bending and C=C stretching vibrations, respectively, of the *trans*-$(CH)_x$ chains, as assigned in the literature [31]. Interestingly, the phonon sidebands in the *cis*-$(CH)_x$ absorption spectrum are dominated by the excited state C=C stretching vibration at ~ 1540 cm$^{-1}$. In contrast, the *trans*-$(CH)_x$ phonon sidebands do not fit any of the resonantly enhanced vibrations of the *trans*-$(CH)_x$ isomer as measured by the RS spectrum (*Fig*. 1d and [31]).

The 80 K EA spectra for the as-synthesized and annealed $(CH)_x$ films are compared with their respective absorption spectra in *Figs*. 2(a) and 2(b), respectively. We note that the EA spectra are free from the assumed background absorption due to light scattering that may be included in the obtained 'absorption' spectrum [24] (See *Fig*. 1(c)). In the as-synthesized film, we observe an oscillatory feature above 1.9 eV that diminishes upon annealing. This supports the assignment that these oscillatory bands with 'zero-crossing' at 2.03, 2.20, and 2.40 eV, respectively (labeled I`, II` and $mA_g^{(c)}$ [22] respectively), correspond to the *cis*-$(CH)_x$ isomer. As for the oscillation with zero-crossing at 1.52, 1.63, 1.72, and 1.82 eV, respectively (labeled I, II, III and IV, respectively) in the as-synthesized film, we note that similar features, albeit red-shifted are also seen in the annealed film with zero crossing at 1.48, 1.60, 1.69, and 1.77 eV, respectively; we thus assign these oscillatory feature as due to the *trans*-(CH)$_x$ isomer in the film.

The EA spectrum of the annealed film also contains a band below the first oscillation feature that appears to be a single peak at 1.32 eV. This band does not have any contribution in the absorption spectrum of the film, and therefore we assign it to an even parity dark (forbidden) exciton, mA$_g$, that is activated by the electric field, which in fact breaks the inversion symmetry in the film [22]. This A$_g$ state is close, but still below the absorption edge; as such, it serves as an 'energy sink' for the photogenerated 1B$_u$ exciton in the *trans*-(CH)$_x$ chains according to Kasha's thermalization rule. This may explain the extremely weak PL emission spectrum of the *trans*-(CH)$_x$ isomer [14], and serves as a straightforward explanation of the nonluminescent nature of this isomer without invoking photogenerated charged soliton pairs using the SSH model [9].

Since the EA spectrum accentuates the oscillatory feature related to the phonon sidebands arising from the vibronic coupling with an excited state vibration, it is easy to see that the energy period, ΔE in the *cis*-(CH)$_x$ part of the spectrum is ~ 185 ± 15 meV (~1480 cm$^{-1}$), which is in fair agreement with the C=C stretching mode frequency (1539 cm$^{-1}$) of this isomer obtained from the RRS spectrum (*Fig*. 1(d)). In contrast, ΔE in the *trans*-(CH)$_x$ part of the EA spectrum is 95 ± 15 meV (~760 cm$^{-1}$) which does not correspond to any of the Raman modes revealed in the RRS spectrum (see *Fig*. 1(d)); in particular it is much lower than the most strongly coupled C=C stretching vibration of the *trans* isomer obtained in the RRS spectrum at 1495 cm$^{-1}$ (*Fig*. 1(d)). We believe that this discrepancy is due to the large attenuation coefficient, γ in the excited state vibrational mode related to the 1B$_u$ exciton in *trans*-(CH)$_x$ due to the fast decay into the mA$_g$ exciton that lies close but below it in this isomer. In this case we may use a simplified 'damped harmonic oscillator' equation [32];

$$(\nu_{ap})^2 = (\nu_0)^2 - \gamma^2 \ , \qquad (1)$$

where $\nu_{ap}$ (= 760 cm$^{-1}$) is the observed or 'apparent' frequency and $\nu_0$ (=1495 cm$^{-1}$) is the unperturbed frequency. Using Eq. (1) we calculate the attenuation γ = 1287 cm$^{-1}$ (or ~39 THz). From the relation γ = τ$^{-1}$ where τ is the 1B$_u$ lifetime, we estimate τ = 26 fs. This short lifetime is in agreement with the 1B$_u$ exciton lifetime estimated directly from the ultrafast transient photoinduced absorption measured in *trans*-(CH)$_x$ [14]. We may also apply Eq.(1) to estimate the 1B$_u$ exciton lifetime in *cis*-(CH)$_x$. From the comparison between the phonon sidebands, namely, $v_{ap}$ (1480 cm$^{-1}$) and the C=C stretching vibration, $v_0$ (=1539 cm$^{-1}$) measured in the RS spectra, $\gamma = 422$ cm$^{-1}$ which leads to $\tau = 78$ fs. This value represents the time during which the photogenerated exciton in *cis*-(CH)$_x$ chains in the film decays into the excited states manifold of the lower energy *trans*-(CH)$_x$ chains and explain the stronger PL emission of this isomer. We note that the two $\tau$ values are given in Table 1.

*Figures* 3(a) and 3(b) show the EA amplitude versus the applied voltage, *V*, that follows a quadratic ($V^2$) response, which is the leading EA term when the material has inversion symmetry, such as the (CH)$_x$ chains. *Figure* 3(c) compares the EA spectrum to that of the absorption derivative spectrum. The EA spectrum of *cis*-(CH)$_x$ follows the Stark shift-related first derivative of the absorption spectrum that contains $v(0-0)$, $v(0-1)$ and $v(0-2)$ phonon sidebands. We note that the absorption derivative spectrum contains a contribution from the scattered light that does not show in the EA spectrum, so the broad EA spectrum can not be compared to the absorption derivative spectrum within the spectral range of this spectral feature [19, 33].

*Figure* 3(d) shows the light polarization dependence of the EA spectrum. Horizontal polarized light is oriented along the direction, **x**, of the applied electric field in the EA measurement. The EA magnitude in the **x** direction is twice larger than that in the **y** direction. This is the consequence of the EA being a nonlinear optical process. In fact, the EA is related to the imaginary part of the 3$^{rd}$ order susceptibility tensor, Im$\chi^{(3)}(0,0,\omega,-\omega)$. Since $\chi^{(3)}$ is a 3$^{rd}$ order tensor, it contains four indexes, namely, $\chi_{ijkl}$. Consequently, the obtained light polarization dependence indicates that $\chi_{xxyy} = \frac{1}{2}\chi_{xxxx}$, which is the case when the (CH)$_x$ fibrils are randomly oriented in the plane of the polyacetylene film [20].

To better understand the EA spectrum in (CH)$_x$ we need to identify its origin. Since the EA spectra should show a Stark shift, a useful way to get the spectrum of the EA source is to integrate the EA spectra of the two isomers. The integrated spectra are shown for the as-synthesized and annealed films in *Figs*. 4(a) and 4(b), respectively, and the resulting peaks are summarized in Table 1. The EA and integrated spectra for different annealing times are shown in the S.I., *Figs*. S3 and S4 respectively, showing that the integrated EA spectrum of the *cis*-(CH)$_x$ resembles its absorption spectrum, except for the band at 2.4 eV, where it is stronger than what would be anticipated for a simple phonon ν(0-2) sideband. We, thus, identify this band in the EA spectrum as due to a superposition of the second phonon sideband of the $1B_u$ exciton at 2.02 eV and an even parity state (namely, $mA_g$) that is activated by the applied field. We note that the energy difference between the $1B_u$ and mAg exciton in *cis*-(CH)$_x$ (~0.4 eV) is similar to many other luminescent PCPs [33,34]. We, therefore, conclude that *cis*-(CH)$_x$ falls in the class of non-degenerate luminescent PCPs.

In contrast, the integrated EA spectrum of the *trans*-(CH)$_x$ isomer in the film (*Fig.* 4(b)) is much narrower than the respective absorption spectrum (see *Fig.* 1(c)). It shows a $1B_u$ band followed by three phonon sidebands. We note that the $1B_u$ energy red-shifts from 1.52 to 1.48 eV upon annealing. This indicates that the *trans*-(CH)$_x$ chains become longer upon isomerization [1, 32, 35]. We can fit the integrated EA spectra of the *trans* and *cis* isomers using the Huang Rhys model:

$$\alpha(\omega) \sim \left[ e^{-S} \sum_{p=0}^{\infty} \frac{S^p}{p!} \frac{1}{(E_{1B_u} + ph\nu - h\omega)} \right] \qquad (2)$$

where $S$ is the Huang Rhys (HR) parameter that determines the relative strength of the successive phonon sidebands and $h\nu$ is the apparent vibration energy. As seen in Table 1, $S$ for the *trans*-(CH)$_x$ isomer is ~ 0.5 that is much larger than $S$ for the *cis* isomer. The relaxation energy, $E_r$ of the $1B_u$ exciton is determined by the $S$ parameter via $E_r = Sh\nu_0$, where $\nu_0$ is the vibration frequency of the most coupled phonon, namely, the C=C stretching mode. Taking the C=C stretching mode energy of 190 meV (1532 cm$^{-1}$), we get $E_r$ = 93 meV (750 cm$^{-1}$). From the EA at room temperature we determined that the $1B_u$ exciton is at $E(1B_u)$=1.53 eV (12,340 cm$^{-1}$). Hence, the photoluminescence band calculated from $E(1B_u)$-$E_r$ = 1.44 eV (11,615 cm$^{-1}$), which matches the PL band measured at room temperature in *trans*-(CH)$_x$ by Sheng *et al.* [14].

The reason that the integrated EA spectrum of *trans*-(CH)$_x$ is much narrower than the absorption spectrum is the chain length dependence of the nonlinear susceptibility in PCPs, namely, $\chi^{(3)}$ ~ $N^3$, where N is the number of C-H units in the polymer chain [19, 22]. Therefore, the EA accentuates the longest chains in the film as is evident from the integrated EA spectrum. This solves the puzzle of the EA spectrum in *trans*-(CH)$_x$ that has remained in the field for about three dozen years [20, 24, 25]. There is no need to relate the EA spectrum to the second derivative of the absorption spectrum as done in a previous study [20]. Hence, the shorter chains in the chain length distribution do not contribute equally to the EA spectrum. This explanation may be also true for the EA spectra of the entire class of the PCPs.

Conclusions

We presented the EA spectra of *cis*- and *trans*-(CH)$_x$ in films of different *cis*-/*trans*-(CH)$_x$ ratio, light polarization and electric field dependence. The EA spectrum of the *cis* isomer can be described by a Stark-shift related first derivative of its absorption spectrum. In contrast to the conclusions of previous works, the EA spectrum of the *trans* isomer can also be described as a first derivative of the absorption spectrum, but of the longest chains in the film. In addition, the EA spectrum of *trans*-(CH)$_x$ reveals a dark mA$_g$ state that lies below the allowed 1B$_u$ exciton. From the comparison of the phonon sidebands revealed in the EA spectrum and the RRS spectrum that are dominated by the C=C stretching vibration we estimate the internal conversion time of 1B$_u$ → mA$_g$ to be of order 30 fs. This explains the non-luminescent character of *trans*-(CH)$_x$ without the need to involve ultrafast charge soliton photogeneration.

Acknowledgements; This work was supported by the NSF grant DMR-2206653. We also acknowledge the Dixon Laser Institute of the University of Utah's Department of Physics & Astronomy for instrumentation support.

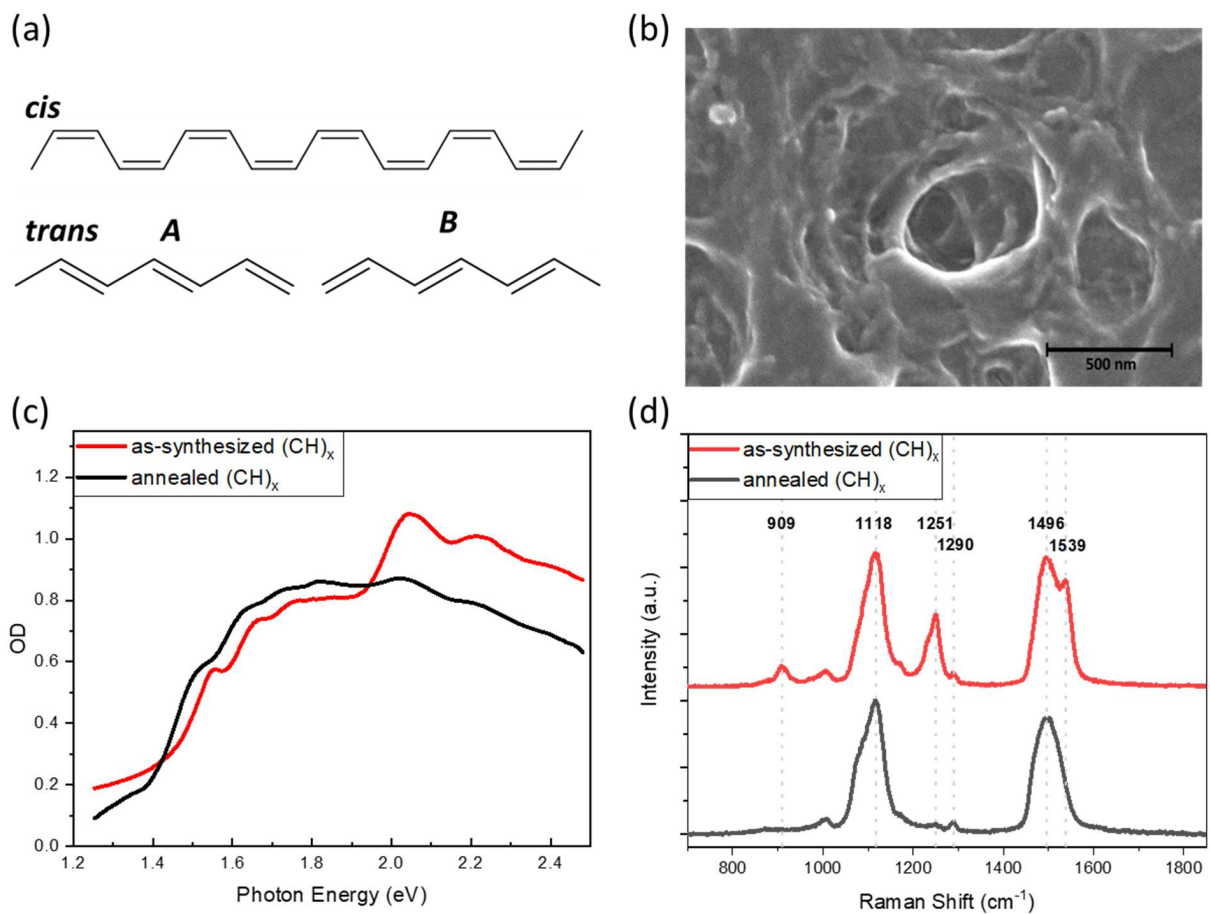

FIG. 1. (a) The *cis*- and *trans*-(CH)$_x$ structures and the two phases of the *trans*-(CH)$_x$ isomer are denoted. (b) SEM image of the fibrils in a (CH)$_x$ thin film. (c) The absorption spectrum of a (CH)$_x$ film before (red line) and after annealing (black line), plotted as optical density (OD). (d) Resonant Raman spectra of a (CH)$_x$ film before (red) and after annealing (black). Few Raman active vibrations are assigned.

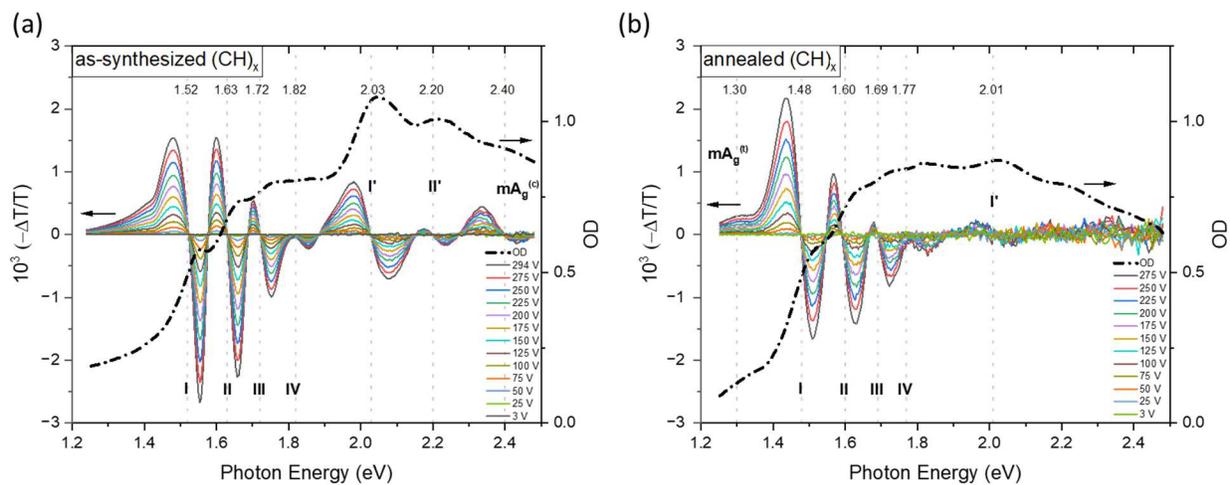

*FIG*. 2. The electroabsorption (EA) spectra for various applied voltages compared with the absorption spectra (OD) (•-•) for 'as synthesized' (a) and 'annealed' (b) $(CH)_x$ thin film. The zero-crossing photon energies, I-IV and the $mA_g^{(t)}$ band are assigned for the *trans*-$(CH)_x$ isomer; and I', II' and $mA_g^{(c)}$ are assigned for the *cis*-$(CH)_x$ isomer.

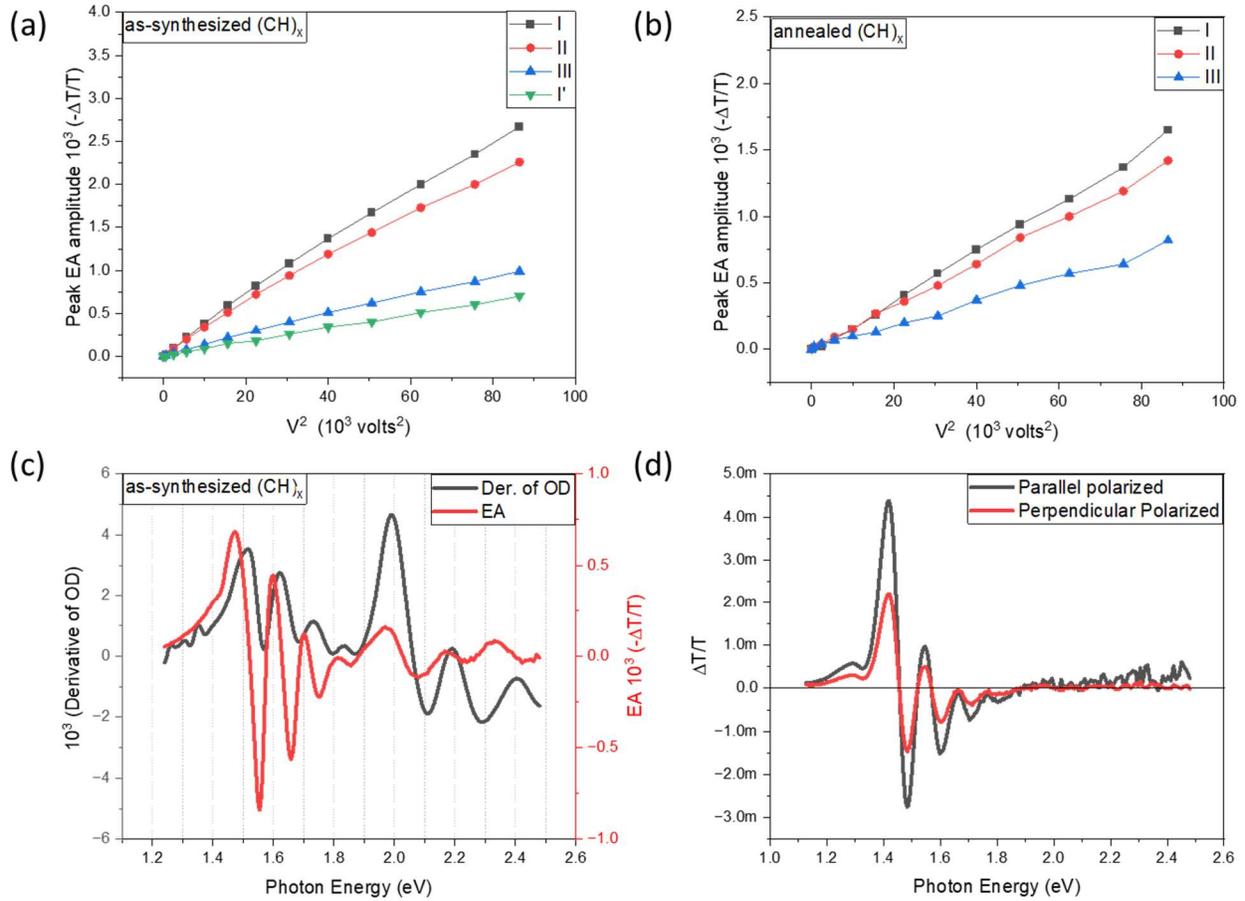

FIG. 3. The EA strengths vs $V^2$ for the as-synthesized (a) and annealed (b) $(CH)_x$ films at several photon energies that correspond to peaks in the EA spectra as denoted in *Fig.* 2(a). (c) Comparison between the EA spectrum and the derivative (Der) of the absorption spectrum for the as-synthesized $(CH)_x$ film. (d) The EA spectra of a $(CH)_x$ film measured at two different light polarization, namely, parallel (black) and perpendicular (red) to the applied electric field.

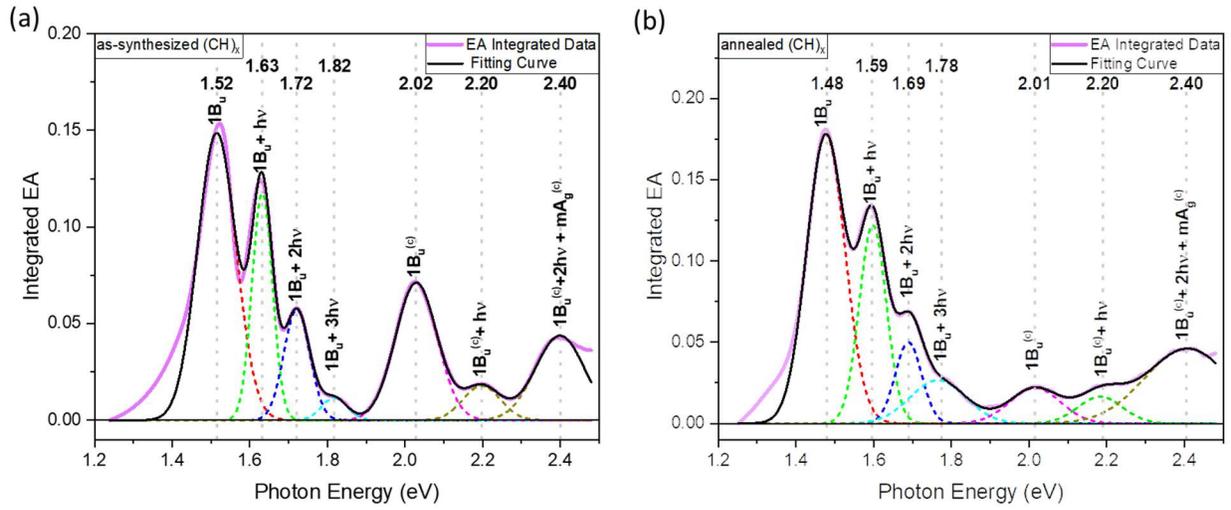

*FIG*. 4. The integrated EA spectra of (a) the as-synthesized and (b) the annealed $(CH)_x$ film. Various bands and phonon replica are assigned (see Table 1). The broken lines through the data are a fit using the Huang-Rhys model (Eq. (2)).

| EA(CH)$_x$ | as-synthesized Peak Center (eV) | annealed Peak Center (eV) |
|---|---|---|
| Peak I | 1.52 | 1.48 |
| Peak II | 1.63 | 1.60 |
| Peak III | 1.72 | 1.69 |
| Peak IV | 1.82 | 1.77 |
| Peak I' | 2.03 | 2.01 |
| Peak II' | 2.20 | 2.18 |
| Peak III' | 2.40 | 2.41 |
| S parameter-*trans* | 0.49 | 0.54 |
| S parameter-*cis* | 0.22 | 0.66 |

Table I. The energies of the peaks in the integrated EA spectra of the as-synthesized and annealed (CH)$_x$ films. Peaks I-IV stand for the *trans*-(CH)$_x$ and peaks I'-III' are for the *cis*-(CH)$_x$. The Huang Rhys parameter, S is also given.

# Supplementary Information (S.I.)

## The influence of dark excitons on the electroabsorption spectrum of polyacetylene

Jaspal Singh Bola[1]*, Ryan M. Stolley[2]*, Prashanna Poudel[1], Joel S. Miller[2], Christoph Boheme[1] and Z. Valy Vardeny[1]**

[1] Department of Physics & Astronomy, University of Utah, Salt Lake City, Utah 84112
[2] Department of Chemistry, University of Utah, Salt Lake City, Utah 84112

**Table of contents**

1. Synthesis of polyacetylene films
2. Electroabsorption (EA) spectroscopy setup
3. Annealing of polyacetylene films

Figure S1. The EA Setup

Figure S2. The EA spectra at 1*f*, 2*f* and 3*f*.

Figure S3. The EA spectra of as-synthesized, 10 min and 30 min annealed $(CH)_x$ films.

Figure S4. Integrated EA spectra for as-synthesized, 10 min and 30 min annealed $(CH)_x$ films.

* These authors contributed equally to this work

** Correspondent author, e-mail val@physics.utah.edu

**S1: Synthesis of polyacetylene films.**

The process used in synthesizing the polyacetylene thin films discussed in this study was a modified version of the method introduced by Shirakawa in 1999 [1]. To conduct the EA measurement, thin films of polyacetylene were deposited onto various substrates with thickness ranging from 0.1-1 μm with the optical density (OD) ~1 at the peak maximum. Making polyacetylene involves using pure acetylene gas and a catalyst for polymerization in a nitrogen-controlled setup. In our study, we made the polyacetylene films at room temperature, resulting in films that had both cis and trans (CH)x isomers.

For the polymerization process, high purity acetylene gas dissolved in acetone underwent purification by passing through subsequent stages of saturated sodium bisulfite solution and concentrated sulfuric acid, respectively.

The cleaned acetylene was subsequently directed through a newly filled glass column (measuring 50 x 3 cm), which contained alternating layers of activated 3 Å molecular sieves and activated alumina (200 mesh). The gas washing system was then fed to a gas manifold (acetylene and vacuum) that had been purged with purified acetylene. For the polymerization catalyst (0.003 M) synthesis, to a dry 100 mL Schlenk flask was sequentially added tetra-n-butoxytitanium(IV), Ti(OBu)4, (251 μL, 0.74 mmol, 1 equiv), purity >97% (Thermo-Fisher), and a triethylaluminium, Et3Al, solution (Sigma Aldrich) in hexanes (2.95 mL, 295 mmol, 4.0 equiv), without further purification and under Schlenk conditions. The resulting solution turned dark brown, and hexanes were evaporated under vacuum.

Freshly distilled cumene (29 mL, Sigma-Aldrich. 98%, distilled from CaH2 under an argon atmosphere) was then added, and the solution was then stirred at room temperature for one hour. The flask was removed from the glove box, attached to a Schlenk line, and heated at 150 °C under nitrogen for one hour. The flask was then cooled to room temperature, sealed, brought into the glove box, and stored at -30 °C. Before use, the catalyst solution was diluted to the desired concentration with dry, degassed toluene.

Thin-film synthesis: The desired polymerization substrate was placed in a pre-dried Schlenk flask equipped with a rubber septum, and then evacuated and backfilled with nitrogen (3x). The apparatus was then equipped with an acetylene manifold, evacuated, and backfilled with acetylene (1 atm). The substrate was then quickly added 2-3 drops of the pre-prepared catalyst solution (0.003 M) via a plastic syringe fitted with an appropriately long 20 gauge needle.

The initially pale brown solution rapidly transitioned to a deep purple solution. To control film thickness, the flask was carefully evacuated within 0.5-3 minutes to prevent bubbling.

Subsequently, the resulting film underwent a 5-minute vacuum treatment to eliminate residual toluene. The deposition apparatus is then brought into a nitrogen-filled glove box under a vacuum. The substrate is removed from the reaction vessel, and the film is then carefully washed (3 x 1 mL) with dry, degassed toluene via a slow stream from a pipette. Finally, the cleaned substrate is then dried under a vacuum for one hour.

**S2: Electroabsorption (EA) spectroscopy setup**

The setup for electroabsorption measurement (see Fig. S1) uses the same setup as that used to measure the optical absorption spectra, with the sole variation being the utilization of a modulated electric field instead of a mechanical chopper. A monochromator equipped with a tungsten lamp is used to send an incident light beam to the sample, and the transmitted light intensity is measured using a silicon detector. The absorption of the material is quantified by measuring the optical density (OD), also referred to as absorbance (A), according to the equation:

$$A = OD = \log(T_0/T)$$

where $T_0$ and $T$ represent the wavelength-dependent transmitted light intensity for the substrate and the substrate with the sample, respectively.

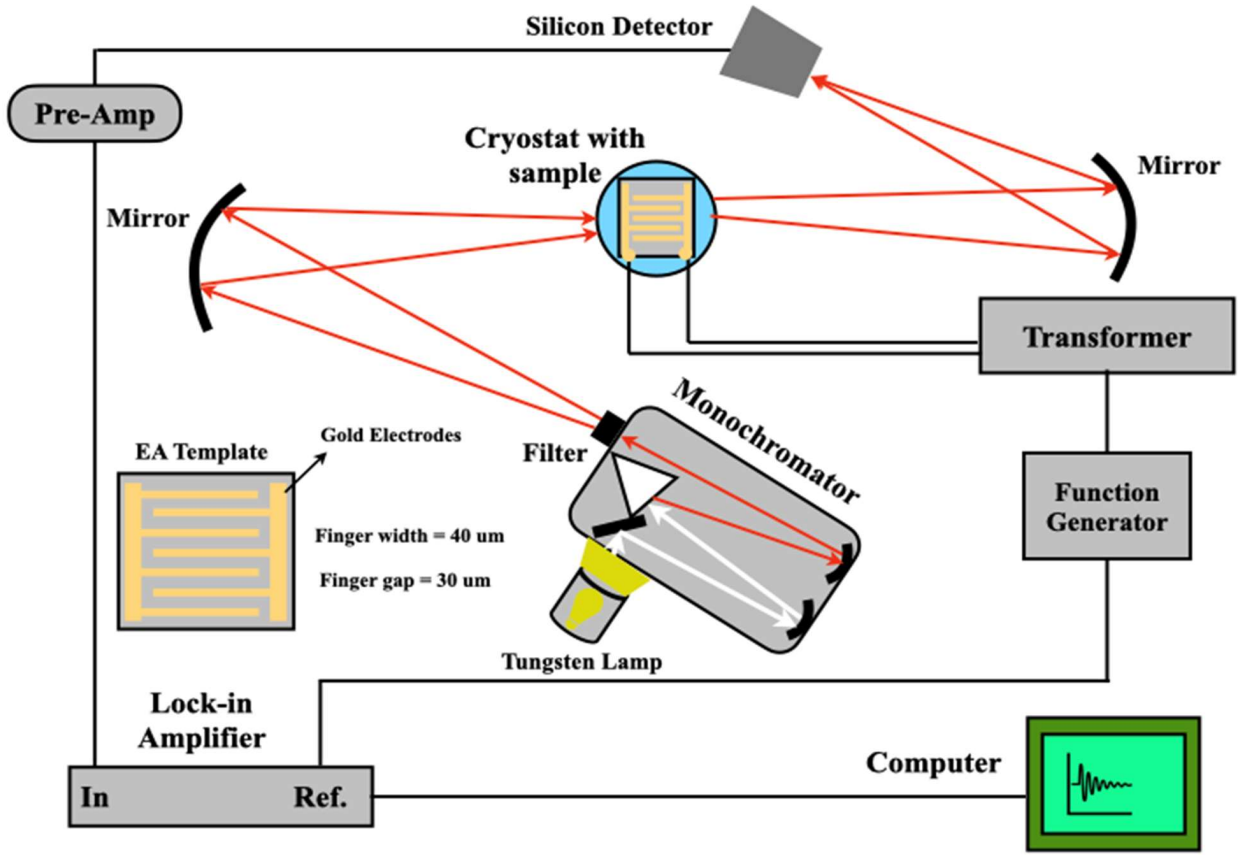

FIG. S1. The electroabsorption setup.

Electroabsorption (EA) measurements were conducted utilizing the configuration depicted in Fig. S1. Enhanced sensitivity of the EA signal was achieved through phase lock-in detection employing electric field modulation. The EA spectrum offers insights into both even and odd parity states. The introduction of the electric field (F) breaks the inversion symmetry, enabling the observation of additional energy states not detectable in linear absorption measurements due to optical selection rules.

The measurement process involves applying an electric field to the sample and measuring the change in transmission, T, of a probe beam from an incandescent lamp. Depending on the wavelength range desired, either a xenon lamp (for UV) or a tungsten lamp (for visible light) is utilized. Specifically for polyacetylene, due to its band edge feature commencing around 1.4 eV, a tungsten lamp is chosen. A sapphire substrate of 12 ×12 mm dimensions is used as a substrate, and 10 nm chrome (chromium??) and 190 nm of gold were RF sputtered on top of it. The interdigitated gold electrode array was patterned using photo-lithography and liftoff technique, maintaining a 30 μm gap between 40 μm fingers. The sample was mounted inside a helium closed cycle cryostat, ensuring proper electrical connections.

In the setup for the EA measurement, electric field substitutes the laser to modulate excited state energies. The probe light beam passes through the film, dispersed through a ¼ met. monochromator and collected by a parabolic mirror before being focused on a UV-enhanced silicon detector. To achieve electric field modulation of about 105V/cm, an AC signal is applied to the sample electrode via a sine wave at a modulation frequency of 1 kHz, amplified to over 300V by a step-up amplifier.

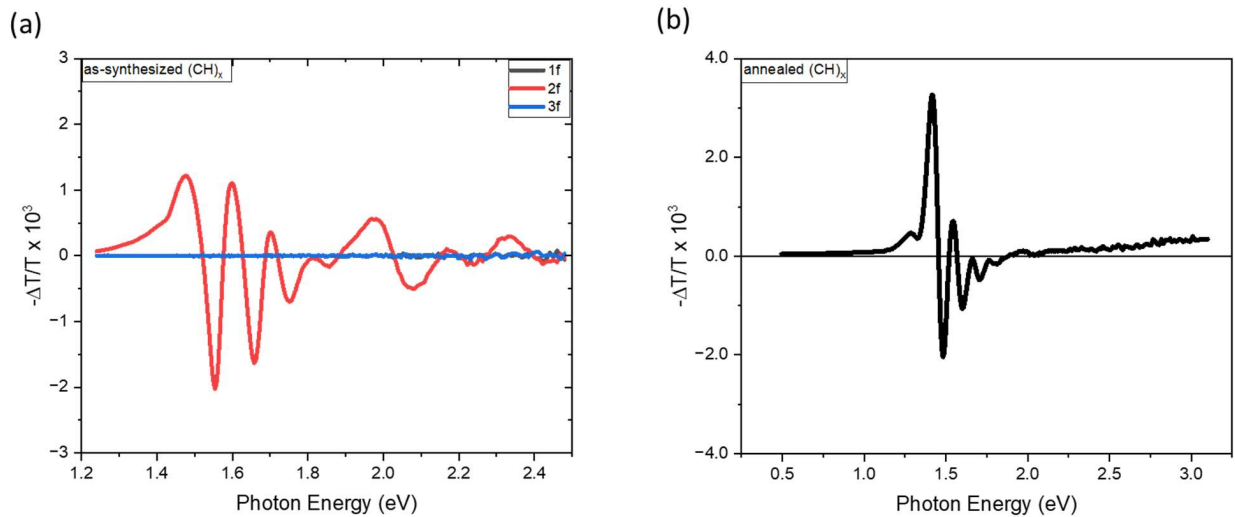

*FIG*. S2. (a) The electroabsorption spectra measured at 1*f*, 2*f* and 3*f* as indicated. (b) The EA spectrum for an annealed $(CH)_x$ film plotted in a broad spectral interval of 0.5 to 3 eV.

The EA signal is detected at 2f due to the mirror symmetry of the conjugated polymer, which responds to a sinusoidal field with frequency f by generating an EA signal observed at 2f. The fundamental frequency is significantly suppressed, and the 2f signal of the EA is proportional to V2. This can be seen in Fig. S2 (a), where 2f of the lock-in amplifier shows the EA signal whereas 1f and 3f do not show any signal. Also, there is no EA signal observed at 2f below the mAg state for trans-$(CH)x$ as shown in Fig. S2(b).

**S3: Annealing of polyacetylene films**

Following the EA measurement, the as-synthesized film was annealed at 150 °C for varying durations in nitrogen environment. Notably, the cis-(CH)x-rich film has a golden shine, , while the trans-(CH)x film appears blue-black in color. Subsequent EA spectrum measurements were conducted after different annealing durations. It was observed that after 30 min of annealing at 150 oC the film was fully converted to trans-(CH)x which can be seen from the disappearance of the EA from the cis-(CH)x at 2 eV. The EA spectra and integrated spectra of as-synthesized, and 10 and 30 min annealed films are shown in Figs. S4 and S5, respectively; these spectra were measured from the same film after different annealing times.

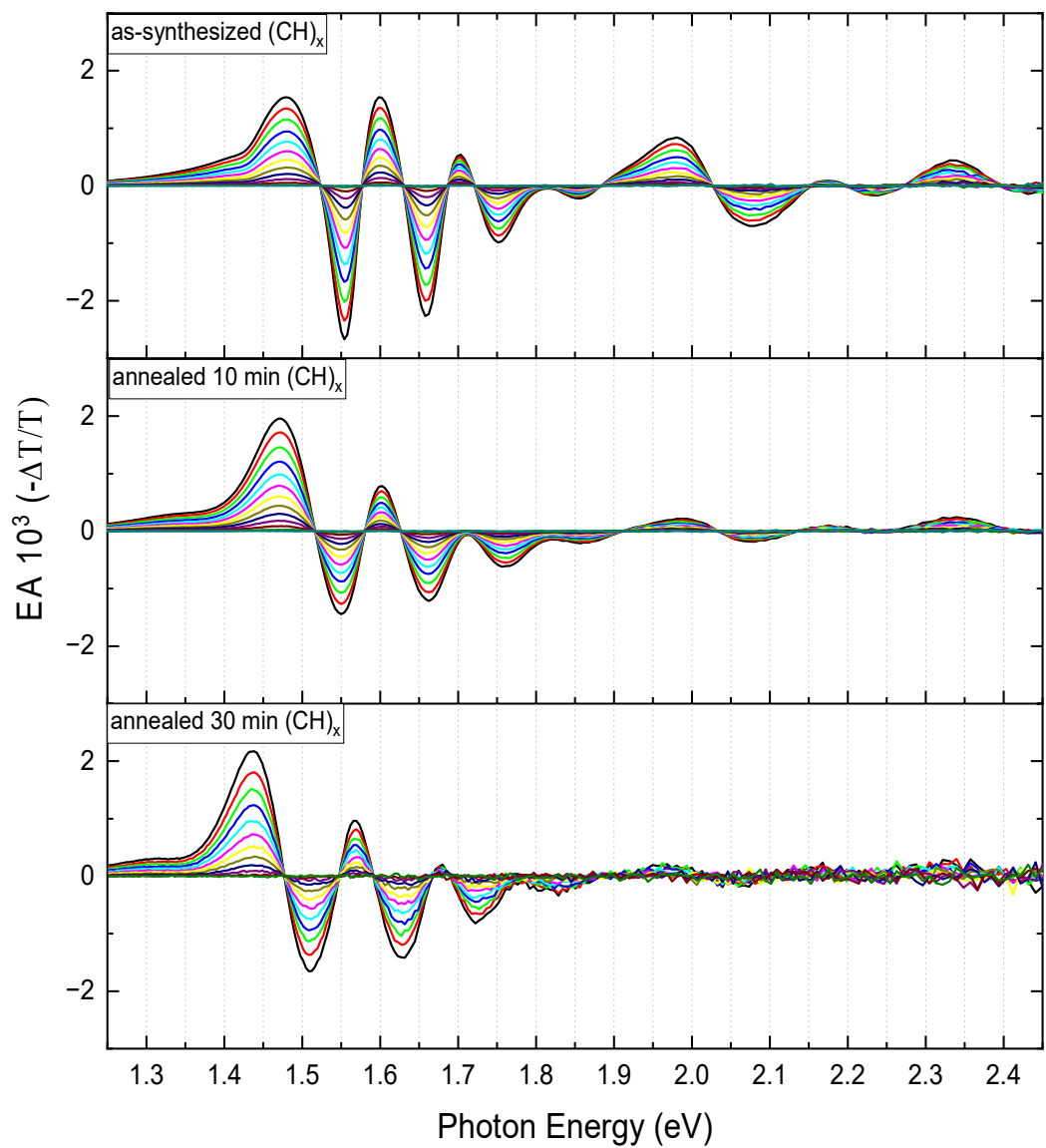

FIG. S4. The EA spectra for as-synthesized, and 10 and 30 min annealed (CH)$_x$ films.

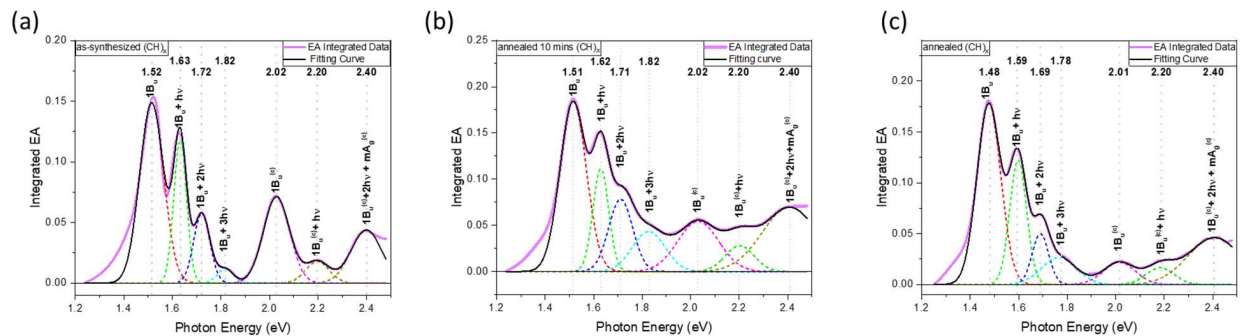

*FIG*. S5. Integrated EA spectra of as-synthesized, 10 and 30 min annealed $(CH)_x$ film.

References

[1] T. S. Liang, K. Akagi, and H. Shirakawa, *Synthesis of Polyacetylene Ultra-Thin Film*, Synthetic Metals **101**, 67 (1999).